\documentclass[10pt]{article}
\usepackage{graphics,fullpage,url,color, multirow}
\usepackage{subfigure}
\usepackage[latin1]{inputenc}
\usepackage[T1]{fontenc} 
\usepackage[english]{babel}
\usepackage{xspace,boxedminipage}
\usepackage{listings}
\usepackage{wrapfig}

\usepackage[small,compact]{titlesec} 
\usepackage[small,it]{caption}

\newcommand{\ie}{\emph{i.e.,} } 
\newcommand{\eg}{\emph{e.g.,} }
\newcommand{\diet}{\textsc{Diet}\xspace}  
\newcommand{\sed}{\textsc{SeD}\xspace}

\newcommand{\ramses}{\textsc{Ramses}\xspace}

\renewcommand{\th}{\textsuperscript{th}\xspace}

\graphicspath{{fig/}}

\title{Cosmological Simulations using Grid Middleware}

\author{\begin{tabular}{ccc}
Y. Caniou, E. Caron, B. Depardon & H. Courtois &  R. Teyssier \\ LIP/ENS-Lyon & CRAL / IFA Hawaii& CEA Saclay \\ {\footnotesize\tt
\{Yves.Caniou,Eddy.Caron,Benjamin.Depardon\}@ens-lyon.fr} & {\footnotesize\tt courtois@ifa.hawaii.edu} & {\footnotesize\tt romain.teyssier@cea.fr}\\
\end{tabular}}

\begin{document}

\maketitle


\section{Introduction} \label{sec:intro}

One way to access the aggregated power of a collection of heterogeneous
machines is to use a grid middleware, such as \diet\cite{CDL+02a},
GridSolve~\cite{GridSolve06} or Ninf~\cite{Ninf99}. It addresses the problem of
monitoring the resources, of handling the submissions of jobs and as an example the inherent
transfer of input and output data, in place of the user.

In this paper we present how to run cosmological simulations using the \ramses
application along with the \diet middleware. We will describe how to write the
corresponding \diet client and server. The remainder of the paper is organized
as follows: Section \ref{sec:diet_overview} presents the \diet middleware.
Section \ref{sec:ramses_overview} describes the \ramses cosmological software
and simulations, and how to interface it with \diet. We show how to write a
client and a server in Section \ref{sec:interfacing}. Finally, Section
\ref{sec:experiments} presents the experiments realized on Grid'5000, the
French Research Grid, and we conclude in Section~\ref{sec:ccl}.

\section{DIET overview}
\label{sec:diet_overview}

\subsection{DIET architecture}

\diet\cite{CDL+02a} is built upon the client/agent/server paradigm. A
\textbf{Client} is an application that uses \diet to solve problems. Different
kinds of clients should be able to connect to \diet: from a web page, a PSE
such as Matlab\footnote{\url{http://www.mathworks.fr/}} or
Scilab\footnote{\url{http://www.scilab.org/}}, or from a program written in C
or Fortran. Computations are done by servers
running a \textbf{Server Daemons ({\sed})}. 
A \sed encapsulates a computational server. 
For instance it can be located on the entry point of a parallel
computer. The information stored by a \sed is a list of the data available on
its server,
all information concerning its load (for example available memory and processor)
and the list of problems that it can solve. 
The latter are declared to its parent agent. The hierarchy of scheduling agents
is made of a \textbf{Master Agent (MA)} and \textbf{Local Agents (LA)} (see
Figure~\ref{fig:archi_appli}).

When a Master Agent receives a computation request from a client,
agents collect computation abilities from servers (through the hierarchy) and
chooses the best one according to some scheduling heuristics. The MA sends back
a reference to the chosen server. A client can be connected to a MA by a
specific name server or by a web page which stores the various MA locations
(and the available problems).
The information stored on an agent is the list of requests, the number of
servers that can solve a given problem and information about the data
distributed in its subtree. For performance reasons, the hierachy of agents
sould be deployed depending on the underlying network topology.




Finally, on the opposite of GridSolve and Ninf which rely on
a classic socket communication layer (nevertheless several problems to this
approach have been pointed out such as the lack of portability or the
limitation of opened sockets), \diet uses Corba. Indeed, distributed object
environments, such as \emph{Java}, \emph{DCOM} or Corba have proven to be a
good base for building applications that manage access to distributed
services. They provide transparent communications in heterogeneous networks,
but they also offer a framework for the large scale deployment of distributed
applications. Moreover, Corba systems provide a remote method invocation
facility with a high level of transparency which does not affect
performance~\cite{DPP01}.

\subsection{How to add a new grid application within \diet ?}

\begin{wrapfigure}{r}{.4\linewidth}
\hspace{.1\linewidth}   
\resizebox{.9\linewidth}{!}{\includegraphics{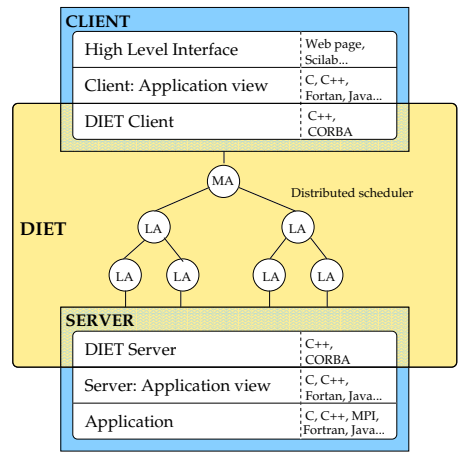}}
\caption{Different interaction layers between \diet core and application view}
\label{fig:archi_appli}
\end{wrapfigure}

The main idea is to provide some integrated level for a grid
application. Figure~\ref{fig:archi_appli} shows these different kinds of level.

The \textbf{application server} must be written to give \diet the ability to
use the application. A simple API is available to easily provide a connection
between the \diet server and the application. The main goals of the {\bf \diet
server} are to answer to monitoring queries from its responsible Local Agent
and launch the resolution of a service, upon an application client request.

The \textbf{application client} is the link between high-level interface and
the \diet client, and
a simple API is provided to easily write one. The main goals of the
\textbf{\diet client} are to submit requests to a scheduler (called Master
Agent) and to receive the identity of the chosen server, and final step, to
send the data to the server for the computing phase.






\section{\ramses overview}
\label{sec:ramses_overview}

\begin{figure}[h]
\begin{center}
\subfigure{
\resizebox{.20\linewidth}{!}{\includegraphics{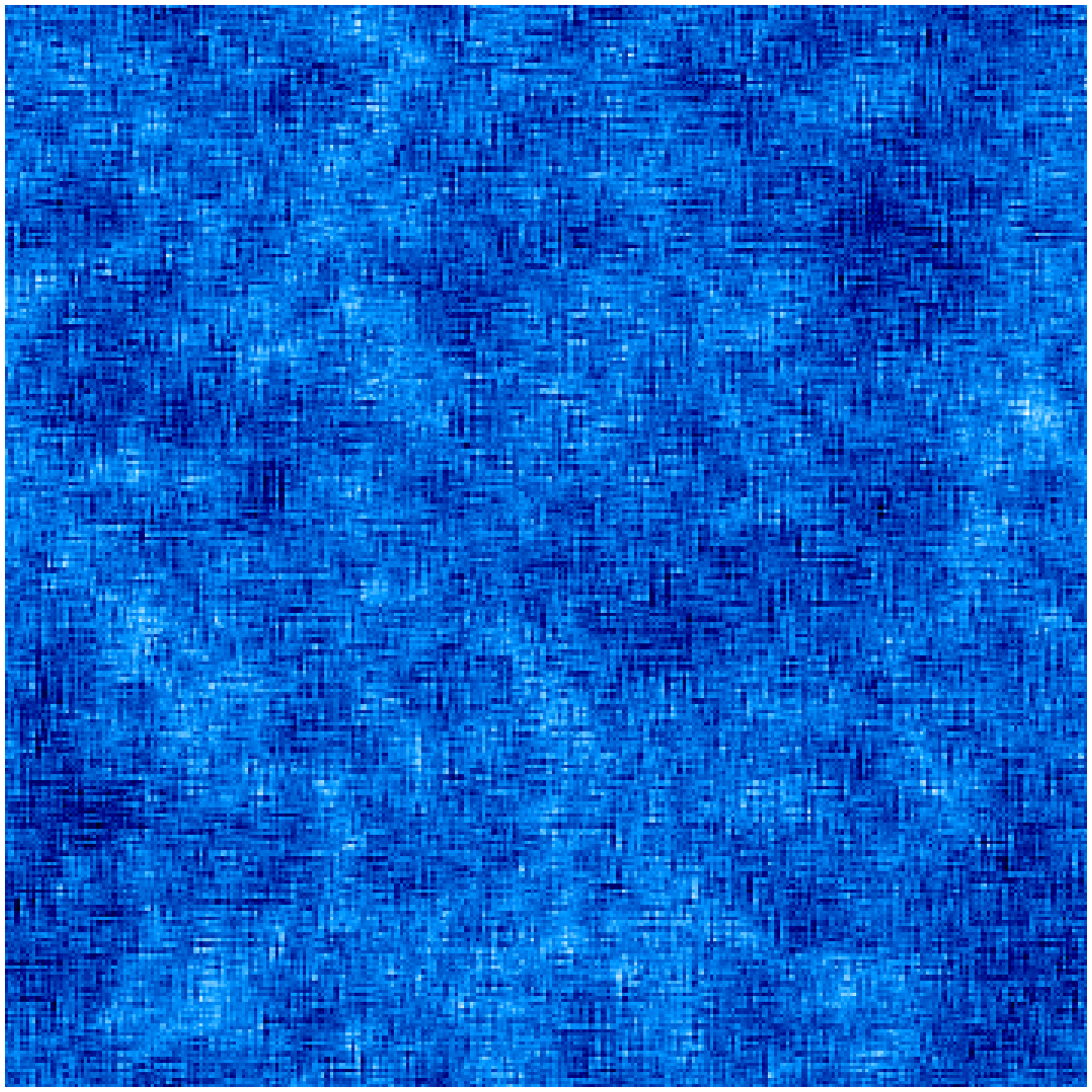}}}
\subfigure{
\resizebox{.20\linewidth}{!}{\includegraphics{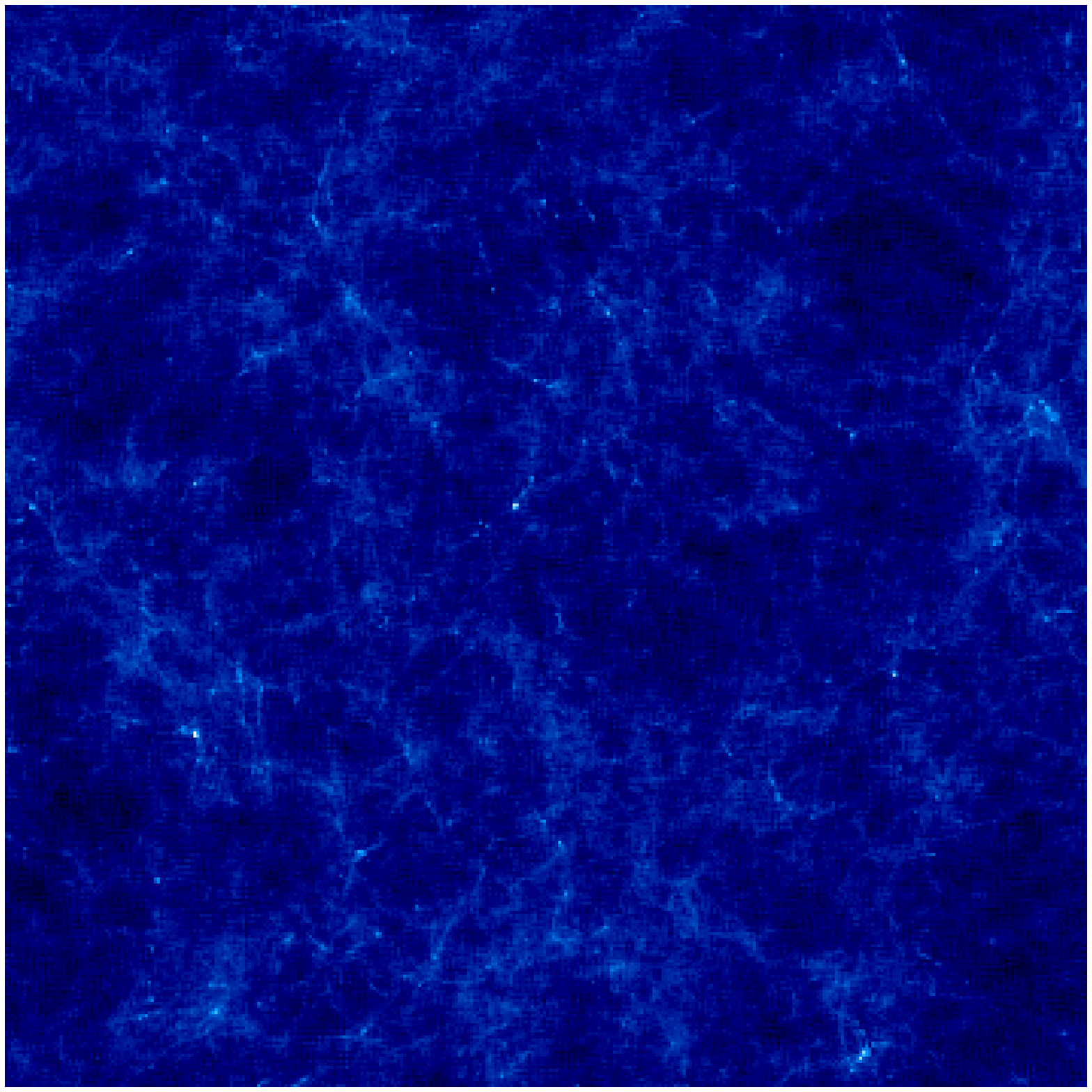}}}
\subfigure{
\resizebox{.20\linewidth}{!}{\includegraphics{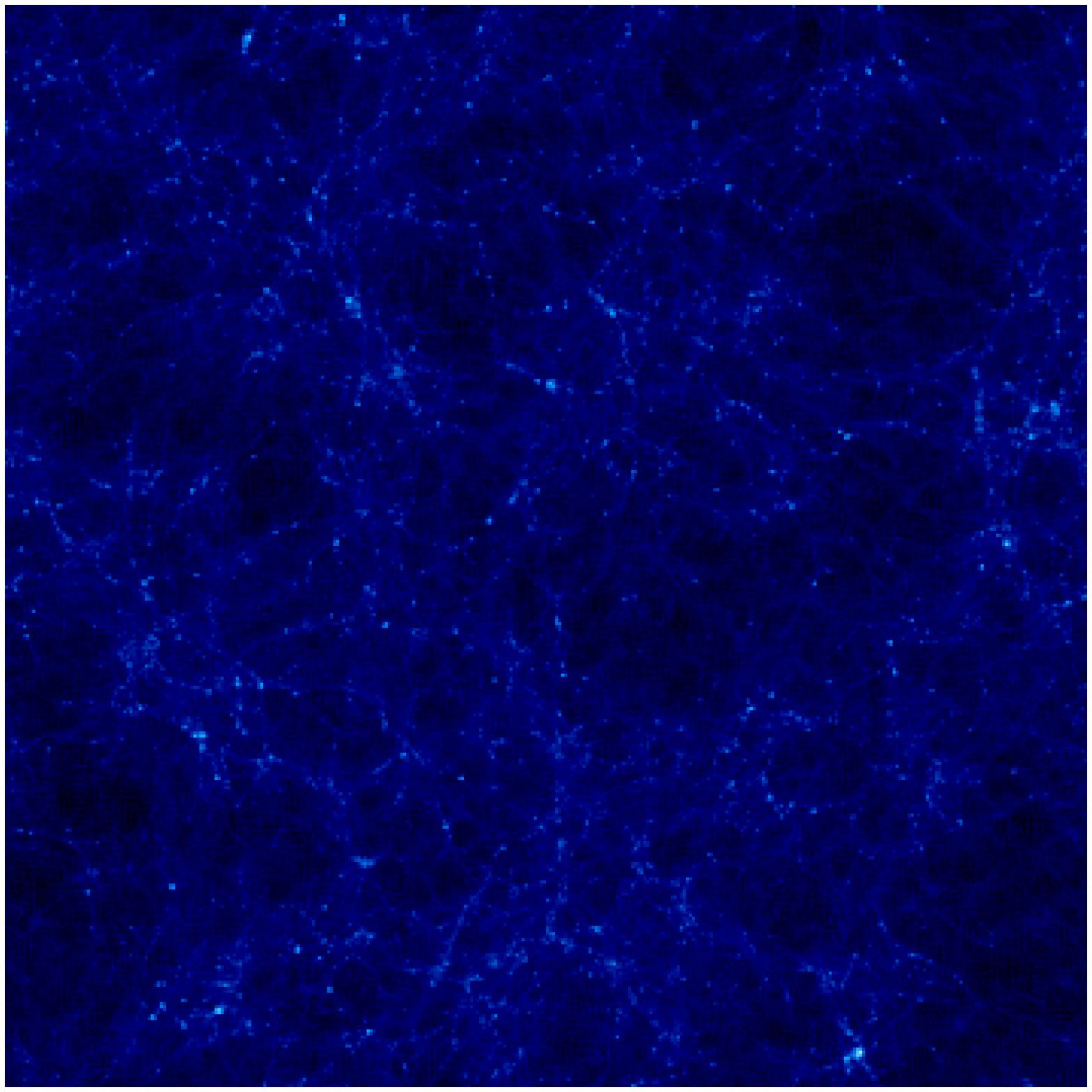}}}
\subfigure{
\resizebox{.20\linewidth}{!}{\includegraphics{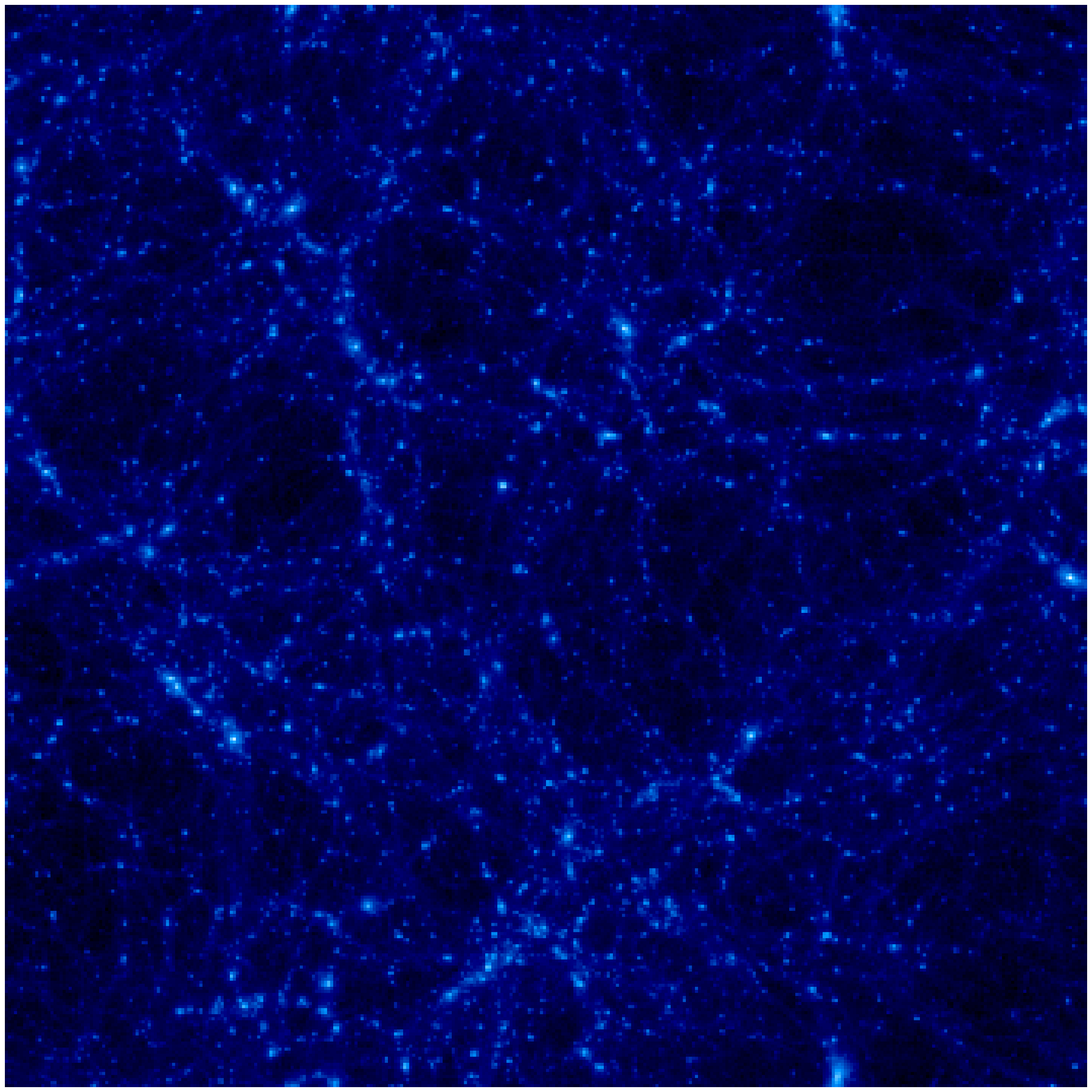}}}
\end{center}
\caption{Time sequence (from left to right) of the projected density field in a
cosmological simulation (large scale periodic box). }
\label{fig:ramses_result}
\end{figure}

\ramses\footnote{\url{http://www.projet-horizon.fr/Codes}} is a typical
computational intensive application used by astrophysicists to study the
formation of galaxies. \ramses is used, among other things, to simulate the
evolution of a collisionless, self-gravitating fluid called ``dark matter''
through cosmic time (see Figure~\ref{fig:ramses_result}). Individual
trajectories of macro-particles are integrated using a state-of-the-art ``N
body solver'', coupled to a finite volume  Euler solver, based on  the Adaptive
Mesh Refinement technics. The computational space is decomposed among the
available processors using a \textit{mesh partitionning} strategy based on the
Peano--Hilbert cell ordering (\cite{Teyssier02, Teyssier06}).

Cosmological simulations are usually divided into two main categories.  Large
scale periodic boxes (see Figure~\ref{fig:ramses_result}) requiring massively
parallel computers are performed on very long elapsed time (usually several
months).  The second category stands for much faster small scale ``zoom
simulations''. One of the particularity of the HORIZON
project
is that it allows the re-simulation of some areas of 
interest for astronomers.

\begin{wrapfigure}{l}{5.5cm} 
\resizebox{1\linewidth}{!}{\includegraphics{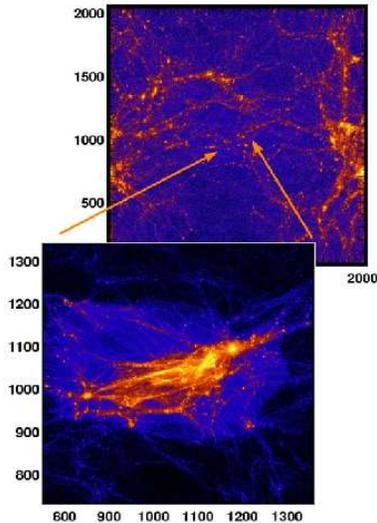}} 
\caption{Re-simulation on a supercluster of galaxies to increase the resolution
\label{fig:zoom}}
\vspace{-.5cm}
\end{wrapfigure}

For example in Figure~\ref{fig:zoom}, a supercluster of galaxies has been
chosen to be re-simulated at a higher resolution (highest number of particules)
taking the initial information and the boundary conditions from the larger box
(of lower resolution). This is the latter category we are interested in.
Performing a zoom simulation requires two steps: the first step consists of
using \ramses on a low resolution set of initial conditions \ie with a small
number of particles) to obtain at the end of the simulation a catalog of ``dark
matter halos'', seen in Figure~\ref{fig:ramses_result} as high-density peaks,
containing each halo position, mass and velocity.  A small region is selected
around each halo of the catalog, for which we can start the second step of the
``zoom'' method. This idea is to resimulate this specific halo at a much better
resolution. For that, we add in the Lagrangian volume of the chosen halo a lot
more particles, in order to obtain more accurate results.  Similar ``zoom
simulations'' are performed in parallel for each entry of the halo catalog and
represent the main resource consuming part of the project.

\ramses  simulations  are  started  from specific  initial  conditions,
containing  the  initial particle  masses,  positions  and velocities.   These
initial  conditions are  read from  Fortran  binary files,  generated using  a
modified      version     of      the      \textsc{Grafic}\footnote{     \url{
http://web.mit.edu/edbert}} code.  This  application generates Gaussian random
fields at  different resolution levels, consistent  with current observational
data obtained  by the WMAP\footnote{\url{http://map.gsfc.nasa.gov}} satellite
observing  the cosmic microwave  background radiation.   Two types  of initial
conditions can be generated with \textsc{Grafic}:

\partopsep -0.15cm
\begin{itemize}
\itemsep -0.15cm
\item  single  level:  this  is  the ``standard''  way  of  generating  initial
conditions. The resulting  files are used to perform  the first, low-resolution
simulation, from which the halo catalog is extracted.
\item  multiple  levels:  this  initial  conditions are  used  for  the  ``zoom
simulation''.  The resulting files consist of multiple, nested boxes of smaller
and smaller  dimensions, as  for Russian dolls.   The smallest box  is centered
around the halo  region, for which we have locally a  very high accuracy thanks
to a much larger number of particles.
\end{itemize}

The result of the simulation is a set of ''snaphots''. Given a list of time
steps (or expansion factor), \ramses outputs the current state of the universe
(\ie the different parameters of each particules) in Fortran binary
files. 
These files need post-processing with \textsc{Galics} softwares: HaloMaker,
TreeMaker and GalaxyMaker. These three  softwares are meant to be used
sequentially, each of them producing different kinds of information:

\begin{itemize}
\itemsep -0.15cm
\item HaloMaker: detects dark matter halos present in \ramses output
files, and creates a catalog of halos
\item TreeMaker: given the catalog of halos, TreeMaker builds a merger tree:
it follows the position, the mass, the velocity of the different particules
present in the halos through cosmic time
\item GalaxyMaker: applies a semi-analytical model to the results of TreeMaker
to form galaxies, and creates a catalog of galaxies
\end{itemize}

\section{Interfacing \ramses within \diet}
\label{sec:interfacing}

\subsection{Architecture of underlying deployment}

The current version of \ramses requires a NFS working directory in order  to
write the output files, 
hence restricting the possible types of solving architectures. Each \diet
server will be in charge of a set of machines (typically 32 machines to run a
$256^3$ particules simulation) belonging to the same cluster. For each
simulation the generation of the initial conditions files, the processing and
the post-processing are done on the same cluster: the server in charge of a
simulation manages the whole process.



\subsection{Server design}

The \diet server is a library. So the \ramses server requires to define the
\texttt{main()} function, which contains the problem profile definition and
registration, and the solving function, whose parameter only consists of the
profile and named after the service name, \texttt{solve\_serviceName}.

The \ramses solving function contains the calls to the different programs used
for the simulation, and which will manage the MPI environment required by
\ramses. It is recorded during the profile registration.

The \sed is launched with a call to \texttt{diet\_SeD()} in the \texttt{main()}
function, which will never return (except if some errors occur). The \sed forks
the solving function when
requested.





\subsubsection{Defining services}
\label{sec:defservice}

To match client requests with server services, clients and servers must use the
same problem description. A unified way to describe problems is to use a name
and define its arguments. The \ramses service is described by a profile
description structure called \texttt{diet\_profile\_desc\_t}. Among its fields,
it contains the name of the service, an array which does not contain data, but
their characteristics, and three integers \texttt{last\_in, last\_inout} and
\texttt{last\_out}. The
structure is defined in \texttt{DIET\_server.h}.

The array is of size $last\_out + 1$. Arguments can be:
\begin{description}
\item{IN:} Data are sent to the server. The memory is allocated by the user.
\item{INOUT:} Data, 
  allocated by the user,
  are sent to the server and brought back into the same memory zone after the
  computation has completed, {\emph without any copy}. Thus freeing this memory
  while the computation is performed on the server would result in a
  segmentation fault when data are brought back onto the client.
\item{OUT:} Data are created on the server and brought back into a newly
  allocated zone on the client. This allocation is performed by \diet. After
  the call has returned, the user can find its result in the zone pointed at by
  the \emph{value} field. Of course, \diet cannot guess how long the user needs
  these data for, so it lets him/her free the memory  
  with \texttt{diet\_free\_data()}.
\end{description}

The fields \emph{last\_in}, \emph{last\_inout} and \emph{last\_out} of the
structure respectively point at the indexes in the 
array of the last IN, last INOUT and last OUT arguments.

Functions to create and destroy such profiles are defined with the prototypes
below. Note that if a server can solve multiple services, each profile should
be allocated.

{\tiny
\begin{lstlisting}[language=C]
diet_profile_desc_t *diet_profile_desc_alloc( const char* path, int last_in, int last_inout, int last_out );
diet_profile_desc_t *diet_profile_desc_alloc( int last_in, int last_inout, int last_out );

int diet_profile_desc_free(diet_profile_desc_t *desc);
\end{lstlisting}
}

The cosmological simulation is divided in two services: \texttt{ramsesZoom1}
and \texttt{ramsesZoom2}, they represent the two parts of the simulation. The
first one is used to determine interesting parts of the universe, while the
second is used to study these parts in details. The \texttt{ramsesZoom2}
service uses nine data. The seven firsts are IN data, and contain the
simulation parameters:
\begin{itemize}
\itemsep -0.15cm
\item a file containing parameters for \ramses
\item resolution of the simulation (number of particules)
\item size of the initial conditions (in $Mpc.h^{-1}$)
\item center's coordinates of the initial conditions (3 coordinates: $c_x$, $c_y$ and $c_z$)
\item number of zoom levels (number of nested boxes)
\end{itemize}
The last two are an integer for error controls, and a file containing the
results obtained from the simulation post-processed with \textsc{Galics}. This
conducts to the following inclusion in the server code (note: the same
allocation must be performed on the client side, with the
\texttt{diet\_profile\_t} structure):
{\tiny
\begin{lstlisting}[language=C]
/* arg.profile is a diet_profile_desc_t * */
arg.profile = diet_profile_desc_alloc("ramsesZoom2", 6, 6, 8);
\end{lstlisting}
}



Every argument of the profile must then be set with
\texttt{diet\_generic\_desc\_set()} defined in \texttt{DIET\_server.h}, like:

{\tiny
\begin{lstlisting}[language=C]
  diet_generic_desc_set(diet_parameter(pb,0), DIET_FILE, DIET_CHAR);
  diet_generic_desc_set(diet_parameter(pb,1), DIET_SCALAR, DIET_INT);
\end{lstlisting}
}

\subsubsection{Registering services}

Every defined service has to be added in the service table before the \sed is
launched. The complete service table API is defined in \texttt{DIET\_server.h}:
{\tiny
\begin{lstlisting}[language=C]
typedef int (* diet_solve_t)(diet_profile_t *);

int diet_service_table_init(int max_size);
int diet_service_table_add(diet_profile_desc_t *profile, NULL, diet_solve_t solve_func);
void diet_print_service_table();
\end{lstlisting}
}

The first parameter, \emph{profile}, is a pointer on the profile previously
described (section~\ref{sec:defservice}). The second parameter concerns the
convertor functionality, but this is out of scope of this paper and never used
for this application. The parameter 
\emph{solve\_func} is the type of the \texttt{solve\_serviceName()} function: a
function pointer used by \diet to launch the computation. Here, the prototype
is then:

{\tiny
\begin{lstlisting}[language=C]
int solve_ramsesZoom2(diet_profile_t* pb)
{
 /* Data downloading */
 /* Computation */
 /* Data uploading */
}
\end{lstlisting}
}

\subsubsection{Data management}

The first part of the solve function (called \texttt{solve\_ramsesZoom2()}) is
to receive data. 
The API provides useful functions to help coding the solve function, \eg get IN
arguments,
set OUT ones, with \texttt{diet\_*\_get()} functions defined in
\texttt{DIET\_data.h}.
Do not forget that the necessary memory space for OUT arguments is allocated by
\diet. So the user should call the \texttt{diet\_*\_get()} functions to
retrieve the pointer to the zone his/her program should write to. To set INOUT
and OUT arguments, one should use the \texttt{diet\_*\_desc\_set()} defined in
\texttt{DIET\_server.h}. These should be called within ``solve'' functions only.


{\tiny
\begin{lstlisting}[language=C]
  diet_file_get(diet_parameter(pb,0), NULL, &arg_size, &nmlPath);
  diet_scalar_get(diet_parameter(pb,1), &resol, NULL);
  diet_scalar_get(diet_parameter(pb,2), &size, NULL);
  diet_scalar_get(diet_parameter(pb,3), &cx, NULL);
  diet_scalar_get(diet_parameter(pb,4), &cy, NULL);
  diet_scalar_get(diet_parameter(pb,5), &cz, NULL);
  diet_scalar_get(diet_parameter(pb,6), &nbBox, NULL);
\end{lstlisting}
}

The results of the simulation are packed into a tarball file if it
succeeded. Thus we need to return this file and an error code to inform the
client whether the file really contains results or not. In the following code,
the \texttt{diet\_file\_set()} function associate the \diet parameter with the
current file. Indeed, the data should be available for \diet, when it sends the
resulting file to the client.

{\tiny
\begin{lstlisting}[language=C]
  char* tgzfile = NULL;
  tgzfile = (char*)malloc(tarfile.length()+1);
  strcpy(tgzfile, tarfile.c_str());
  diet_file_set(diet_parameter(pb,7), DIET_VOLATILE, tgzfile);
\end{lstlisting}
}

\subsection{Client}
\label{sec:client}

In the \diet architecture, a client is an application which uses \diet to 
request a service. The goal of the client is to connect to a Master Agent in
order to 
dispose of a \sed which will be able to solve the problem. Then the client
sends input data to the chosen \sed and, after the end of computation, retrieve
output data from the \sed. \diet provides libraries containing functions to
easily and transparently access the \diet platform.

\subsubsection{Structure of a client program}
\label{sec:cl_struct}

Since the client side of \diet is a library, a client program has to define the
\texttt{main()} function: it uses \diet through function calls.
{\tiny
\begin{lstlisting}[language=C]
#include "DIET_client.h"

int main(int argc, char *argv[])
{
  diet_initialize(configuration_file, argc, argv);
  // Successive DIET calls ...
  diet_finalize();
}
\end{lstlisting}
}

The client program must open its \diet session with a call to
\texttt{diet\_initialize()}. It parses the configuration file given as the
first argument, to set all options and get a reference to the \diet Master
Agent. The session is closed with a call to \texttt{diet\_finalize()}. It frees
all resources, if any, associated with this session on the client, servers, and
agents, but not the memory allocated for all INOUT and OUT arguments brought
back onto the client during the session. Hence, the user can still access them
(and still has to free them !).

The client API follows the GridRPC definition \cite{gridRPC:02}: all
\texttt{diet\_} functions are ``duplicated'' with \texttt{grpc\_} functions.
Both \texttt{diet\_initialize()}/\texttt{grpc\_initialize()} and
\texttt{diet\_finalize()}/\texttt{grpc\_finalize()} belong to the GridRPC API.
 
A problem is managed through a \emph{function\_handle}, that associates a
server to a service name. 
The returned \emph{function\_handle} is associated to the problem description,
its profile, during the call to \texttt{diet\_call()}.

\subsubsection{Data management}

The API to the \diet data structures consists of modifier and accessor
functions only: no allocation function is required, since
\texttt{diet\_profile\_alloc()} allocates all necessary memory for all argument
\textbf{descriptions}. This avoids the temptation for the user to allocate the
memory for these data structures twice (which would lead to \diet errors while
reading profile arguments).

Moreover, the user should know that arguments of the \texttt{\_set} functions
that are passed by pointers are \textbf{not} copied, in order to save memory.
Thus, the user keeps ownership of the memory zones pointed at by these
pointers, and he/she must be very careful not to alter it during a call to
\diet. An example of prototypes:

{\tiny
\begin{lstlisting}[language=C]
int diet_scalar_set(diet_arg_t* arg, void* value, diet_persistence_mode_t mode, diet_base_type_t base_type);
int diet_file_set(diet_arg_t* arg, diet_persistence_mode_t mode, char* path);
\end{lstlisting}
}

Hence the arguments used in the \texttt{ramsesZoom2} simulation are declared as
follows: {\tiny
\begin{lstlisting}[language=C++]
// IN parameters
if (diet_file_set(diet_parameter(arg.profile,0), DIET_VOLATILE, namelist)) {
  cerr << "diet_file_set error on the <namelist.nml> file" << endl;
  return 1;
}
diet_scalar_set(diet_parameter(arg.profile,1), &resol, DIET_VOLATILE, DIET_INT);
diet_scalar_set(diet_parameter(arg.profile,2), &size, DIET_VOLATILE, DIET_INT);
diet_scalar_set(diet_parameter(arg.profile,3), &arg.cx, DIET_VOLATILE, DIET_INT);
diet_scalar_set(diet_parameter(arg.profile,4), &arg.cy, DIET_VOLATILE, DIET_INT);
diet_scalar_set(diet_parameter(arg.profile,5), &arg.cz, DIET_VOLATILE, DIET_INT);
diet_scalar_set(diet_parameter(arg.profile,6), &arg.nbBox, DIET_VOLATILE, DIET_INT);
// OUT parameters
diet_scalar_set(diet_parameter(arg.profile,8), NULL, DIET_VOLATILE, DIET_INT);
if (diet_file_set(diet_parameter(arg.profile,7), DIET_VOLATILE, NULL)) {
  cerr << "diet_file_set error on the OUT file" << endl;
  return 1;
}
\end{lstlisting}
}

It is to be noticed that the OUT arguments should be declared even if their
values is set to NULL. Their values will be set by the server that will execute
the request.

Once the call to \diet is done, we need to access the OUT data. The 8\th
parameter is a file and the 9\th is an integer containing the error code of the
simulation (\texttt{0} if the simulation succeeded):

{\tiny
\begin{lstlisting}[language=C]
int* returnedValue;
size_t tgzSize = 0;
char* tgzPath = NULL;
diet_scalar_get(diet_parameter(simusZ2[reqID].profile,8), &returnedValue, NULL);
if (!*returnedValue) {
  diet_file_get(diet_parameter(simusZ2[reqID].profile,7), NULL, &tgzSize, &tgzPath);
}
\end{lstlisting}
}

\section{Experiments}
\label{sec:experiments}

\subsection{Experiments description}
\label{sec:platform_desc}

Grid'5000\footnote{\url{http://www.grid5000.fr}} is the French Research
Grid. It is composed of 9 sites 
spread all over France, each with 100 to 
1000 PCs, connected by the RENATER Education and Research Network (1Gb/s or
10Gb/s). For our experiments, we deployed a \diet platform on 5 sites (6
clusters).


\begin{itemize}
\itemsep -0.15cm
\item 1 MA deployed on a single node, along with omniORB, the monitoring tools,
and the client
\item 6 LA: one per cluster (2 in Lyon, and 1 in  Lille, Nancy, Toulouse and
Sophia)
\item 11 {\sed}s: two per cluster (one cluster of Lyon had only one \sed due to
reservation restrictions), each controling 16 machines (AMD Opterons
246, 248, 250, 252 and 275)
\end{itemize}

We studied the possibility of computing a lot of low-resolution
simulations. The client requests a $128^3$ particles $100Mpc.h^{-1}$ simulation
(first part). When he receives the results, he requests simultaneously 100
sub-simulations (second part). As each server cannot compute more than one
simulation at the same time, we won't be able to have more than 11 parallel
computations at the same time.

\subsection{Results}

\begin{figure}[t]
\begin{center}
  \resizebox{0.5\linewidth}{!}{\includegraphics{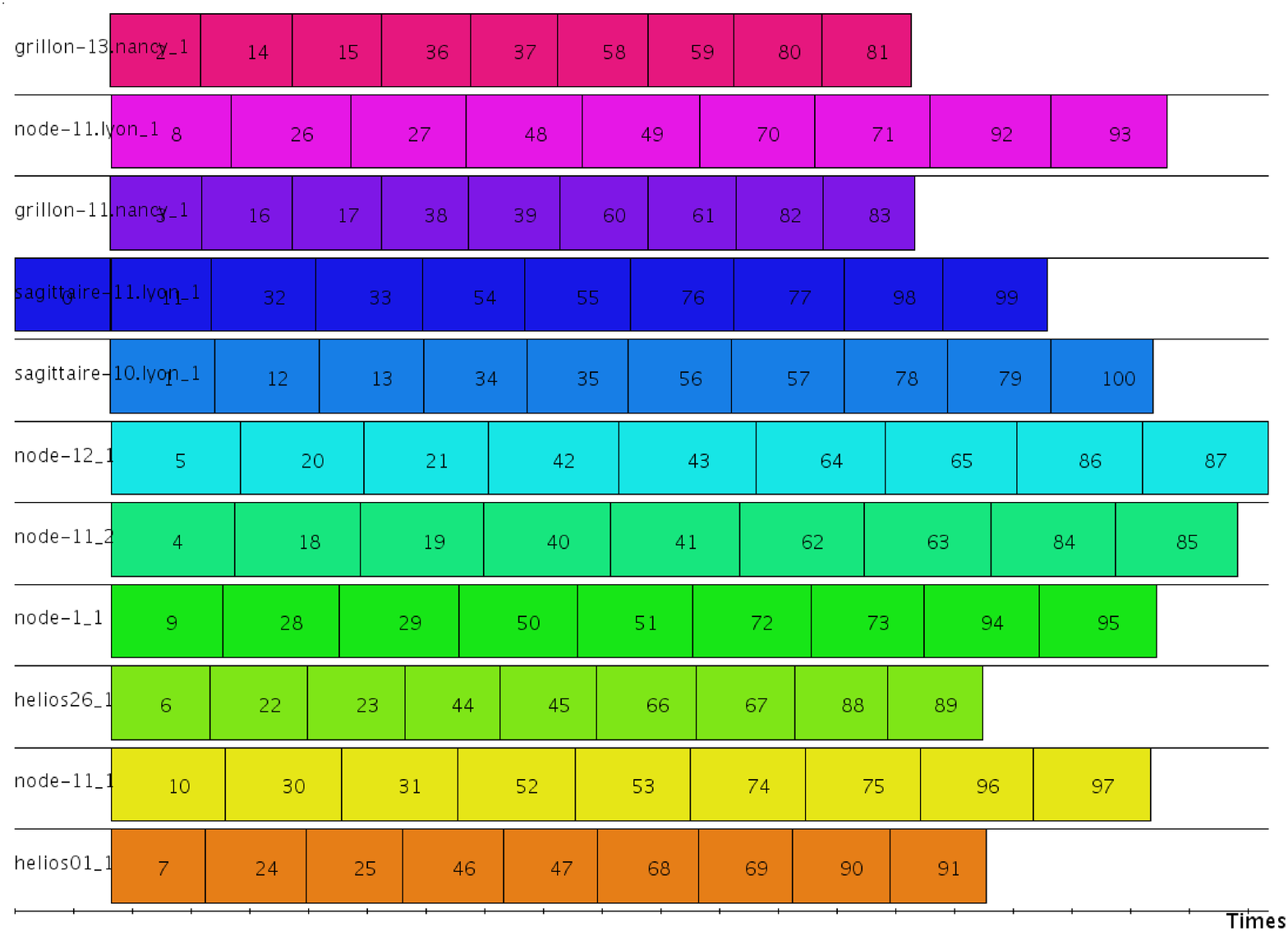}}
  \hspace{.1\linewidth}
  \resizebox{0.5\linewidth}{!}{\includegraphics{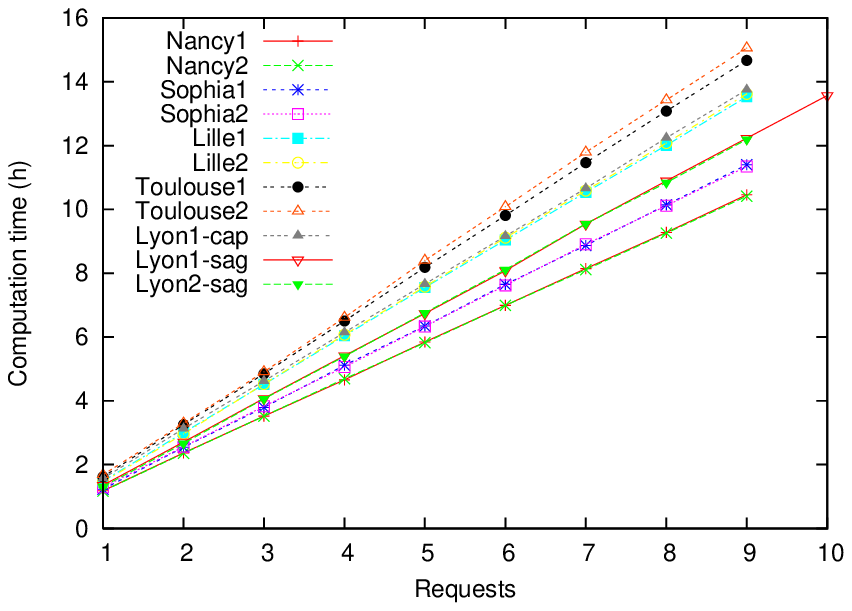}}
\end{center}
\caption{Simulation's distribution over the {\sed}s: on the left, the Gantt chart; on the right, the execution time of the 100 sub-simulations for each \sed}
\label{fig:comp_time_expe}
\end{figure}

The experiment (including both the first and the second part of the simulation)
lasted 16h 18min 43s (1h 15min 11s for the first part and an average of 1h
24min 1s for the second part).  After the first part of the simulation, each
\sed received 9 requests (one of them received 10 requests) to compute the
second part (see Figure~\ref{fig:comp_time_expe}, left). As shown in
Figure~\ref{fig:comp_time_expe} (right) the total execution time for each \sed
is not the same : about 15h for Toulouse and 10h30 for Nancy. Consequently, the
schedule is not optimal. The equal distribution of the requests does not take
into account the machines processing power.  In fact, at the time when \diet
receives the requests (all at the same time) the second part of the simulation
has never been executed, hence \diet doesn't know anything on its processing
time, the best it can do is to share the total amount of requests on the
available {\sed}s. A better  makespan could be attained by writing a plug-in
scheduler\cite{CCDS06}.

\begin{figure}[t]
\begin{center}
\resizebox{0.6\linewidth}{!}{\includegraphics{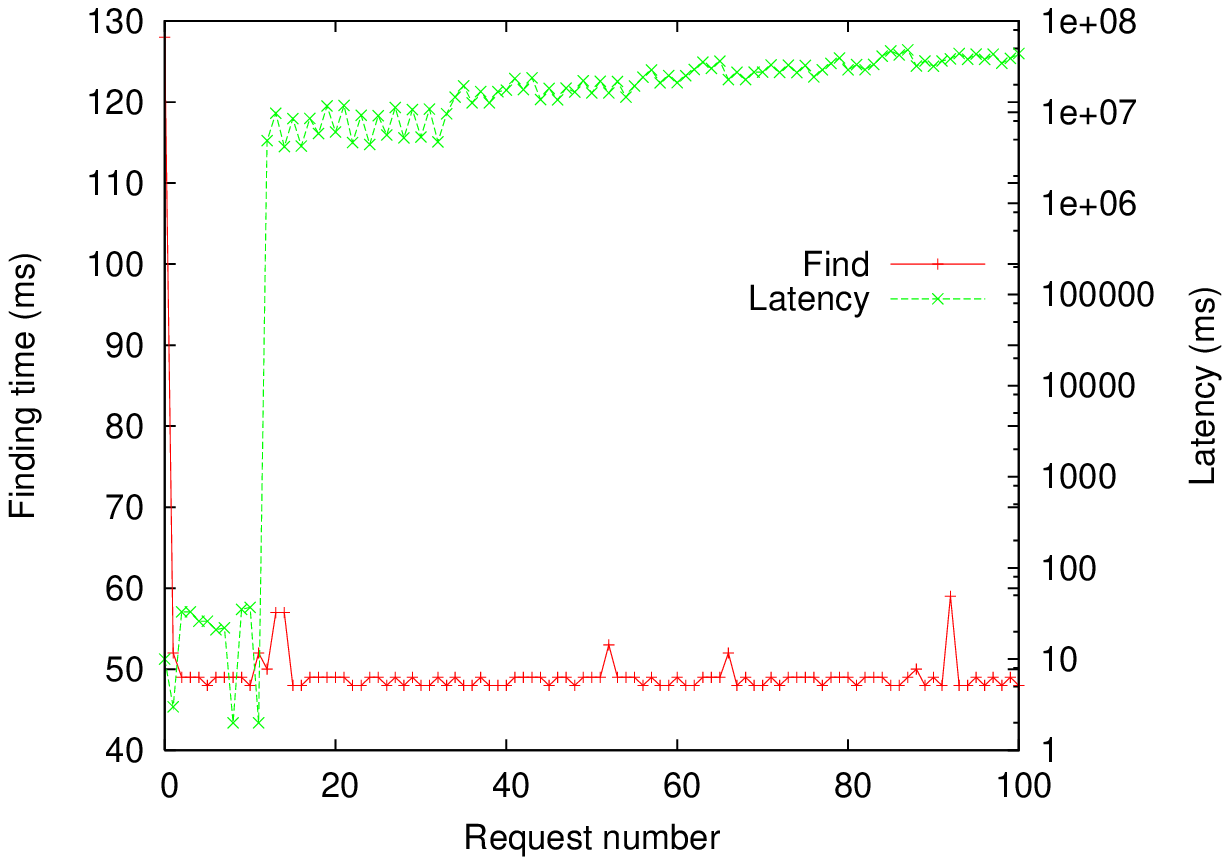}}
\end{center}
\caption{Finding time and latency}
\label{fig:lat_find_expe}
\end{figure}

The benefit of running the simulation in parallel on different clusters is
clearly visible: it would take more than 141h to run the 101 simulation
sequentially.  
Furthermore, the overhead induced by the use of \diet 
is extremely low. Figure~\ref{fig:lat_find_expe} shows the time needed to find
a suitable \sed for each
request, 
as well as in log scale, the latency (\ie the time needed to send the data from
the client to the chosen \sed, plus the time needed to initiate the service).

The finding time
is low and nearly constant (49.8ms on average). 
The latency grows rapidly. Indeed, the client requests 100 sub-simulations
simultaneously, and each \sed cannot compute more than one of them at the same
time. Requests cannot be proceeded until the completion of the precedent
one. This waiting time is taken into account in the latency. Note that the
average time for initiating the service is 20.8ms (taken on the 12 firsts
executions). The average overhead for one simulation is about 70.6ms, inducing
a total overhead for the 101 simulations of 7s, which is neglectible compared
to the total processing time of the simulations.


\section{Conclusion}\label{sec:ccl}

In this paper, we presented the design of a \diet client and server based on
the example of cosmological simulations. As shown by the experiments, \diet is
capable of handling long cosmological parallel simulations: mapping them on
parallel resources of a grid, executing and processing communication transfers.
The overhead induced by the use of \diet is neglectible compared to the
execution time of the services. Thus \diet permits to explore new research
axes in cosmological simulations (on various low resolutions initial
conditions), with transparent access to the services and the data.

\small
\bibliographystyle{plain} 
\bibliography{diet_ramses}
\end{document}